\documentstyle[preprint,aps]{revtex}

\begin{document}  

\preprint{\vbox{\hbox{DART-HEP-95/06}
\hbox{hep-ph/9511295}}}

\title{On The Absence Of Open Strings In A Lattice-Free
Simulation Of Cosmic String Formation}

\author{Julian Borrill}

\address{Blackett Laboratory, Imperial College, London SW7 2BZ,
U.~K.\\
and\\
Department of Physics and Astronomy, Dartmouth College, Hanover,
NH 03755, U.~S.~A.\footnote{Present address}}

\date{\today} 

\maketitle
\tighten

\begin{abstract}
Lattice-based string formation algorithms can, at least in
principle, be reduced to the study of the statistics of the
corresponding aperiodic random walk. Since in three or more
dimensions such walks are transient this approach necessarily
generates a population of open strings. To investigate whether
open strings are an artefact of the lattice we develop an
alternative lattice-free simulation of string formation.
Replacing the lattice with a graph generated by a minimal
dynamical model of a first order phase transition we obtain
results consistent with the hypothesis that the energy density
in string is due to a scale-invariant Brownian distribution of
closed loops alone.

\end{abstract}

\narrowtext

\section{Introduction}

Cosmic strings are the longest standing candidate for a source
of primordial cosmological perturbations arising from particle
physics beyond the standard model. First described by Kibble
almost 2 decades ago \cite{K} they have yet to be demonstrated
to be incompatible with any of the subsequent cosmological
observations, from the anisotropies in the cosmic microwave
background to the distribution of luminous matter on the largest
scales. The field theoretic requirements of the basic model are
simple; the spontaneous breaking of some symmetry of our theory
yielding a degenerate vacuum manifold whose first homotopy group
is non-trivial. The single free parameter in the theory is then
the energy scale of the, typically GUT, symmetry breaking. The
complexity of the ensuing cosmology (in part responsible for the
model's longevity) lies in the need to track the non-linear
evolution of the string-bearing field through the symmetry
breaking phase transition and on to the epoch of any observable
consequences.

The starting point for any analysis of string cosmology is in
the statistics of their initial distribution. Unless terminating
at a pair of monopoles or spatial singularities, all strings
must be either closed loops or open (and hence infinite). Here
we are interested in the initial distribution of strings both
between and within each of these populations. We can trivially
define a fraction $0 \leq f_{\rm c} \leq 1$ of the string energy
density to be in closed loops, leaving $f_{\rm o} = 1 - f_{\rm c}$ in
open string; these are the quantity of most interest in this
Letter. Further, it is easy to show, under the assumption that
the string loops are Brownian and scale-invariant, that the
number density of loops with lengths in the range $[l, l+\delta
l)$ must go as \cite{VV}
\begin{equation}
\label{eLD}
\delta n \propto l^{-5/2} \; \delta l .
\end{equation}

Numerical simulations of string formation assign phases taken
from the $S_{\rm 1}$ manifold of vacuum states to the vertices
of some 3-dimensional lattice\footnote{Note that by `lattice' we
mean a regular array of connected vertices, whilst the oxymoron
`random lattice' is termed a graph.}. Along the lattice edges
the field is taken to follow the shortest path on the manifold
--- the so-called geodesic rule. Each face of the lattice is
then inspected to determine whether, in traversing its boundary,
the field covers the entire vacuum manifold. If so then a zero
of the field is necessarily present somewhere within the face,
and a string segment is located passing though it. Such a
segment connects the centres of the adjacent lattice volume
elements associated with the face, and taken together all the
segments join to form a population of complete connected
strings. If the lattice is periodic then all the strings are
closed loops; otherwise strings which intersect the lattice
boundaries simply terminate there.

The original simulation \cite{VV} took phases at random from the
minimal 3-point discretisation of $S_{\rm 1}$ on a cubic
lattice, resolving the ambiguity when 4 string segments met in a
single cube by chance. Later refinements to this approach
included the use of a tetrahedral lattice \cite{SF,BSD,HS}, a
continous vacuum manifold \cite{SF,LP}, and a dynamical
allocation of the field phases \cite{H,Sh}. However, the key
common feature of all these simulations is that the field is set
at the vertices of a lattice, and hence that the strings form
along the edges of its dual. In every case such simulations do
indeed generate strings which are Brownian and scale invariant,
with the majority of the string energy density being in open
strings. The exact fraction depends on the details of the
simulation, and in particular on the geometry of the lattice and
its dual. For example, the simple cubic lattice with its simple
cubic dual gives $f_{\rm o} \sim 0.8$ \cite{VV}, whilst the 
tetrahedral lattice with tetrakaidekahedral dual gives $f_{\rm o}
\sim 0.63$ \cite{BSD,HS}.

Since such simulations construct strings as sequences of
vertices on a lattice it is not surprising that they can also be
represented as random walks. Since there is an excluded volume
around the path of a single string, giving a directional bias
away from its origin, we might naively expect the ensemble to
have the statistics of a self-avoiding random walk. However, if
the strings are dense the associated ensemble of excluded
volumes removes this bias, and the statistics turn out to be
those of a Brownian random walk instead. Moreover it is known
that in 3 or more dimensions an aperiodic random walk on a
lattice cannot be recurrent \cite{Sp}, and so has a non-zero
probability of not returning to the origin, giving a non-zero
fraction of open string. Exploiting this representation,
analyses of the statistics of Brownian random walks on the
appropriate lattices give similar results, with $f_{\rm o} \geq 0.6$
\cite{SF,ACR}.

Since open strings appear to be an inevitable consequence of the
lattice-based formation algorithms it is desirable to consider
formulations which do not involve a lattice, and use them test
the alternative hypothesis that the string energy density may be
accounted for solely in terms of a scale-invariant Brownian
population of closed loops. The results of one such algorithm
are presented in this Letter.

\section{Algorithm}

The use of a lattice in string formation simulations allows us
to know in advance where the strings may be found, and hence to
bypass the dynamics of the phase transition. Without a lattice
we must therefore include the dynamics explicitly, and for
computational tractability we adopt the simplest possible
dynamical model of a first order phase transition. Spherically
symmetric bubbles of the true vacuum (with phases chosen at
random from the minimally discretised manifold) are nucleated at
random space-time events in a periodic false vacuum
background. They are then taken to expand uniformly at the speed
of light for the duration of the simulation, colliding with one
another, trapping regions of false vacuum, and generating
strings.

The exterior (ie. outside of any other bubble, and so in the
false vacuum) intersection of the surfaces of any 3 bubbles of
differing phases generates a string segment with open ends. As
the bubbles continue to expand this segment traces the locus of
intersection of their surfaces, lengthening and, except in the
case where the 3 bubbles are the same size, bending (cf.
lattice-based simulations where the bubbles are always the same
size, and the string segments are necessarily straight). As the
bubbles continue to expand they eventually meet a fourth
bubble. One and only one of the collisions between this bubble
and each pair of the original triplet also generates a string
segment. The final 4 bubble intersection event then sees the
joining of one of the ends of each segment to form a new
extended segment. The exterior intersection of the surfaces of
any 4 bubbles can therefore also be viewed as the meeting of 4
3-bubble intersection loci, and hence of the joining of either 0
or 2 string segments.

The key to this algorithm is the location of the points of
intersection of all bubble triplets and quadruplets. However
since the bubbles are spheres of known centres and predictably
time-dependent radii these points of intersection can be
determined analytically. Having calculated the location of all
the 3- and 4-bubble intersections in the simulation space-time
we can then construct the corresponding string distribution and
determine its statistics. Taking the 4-bubble intersection
points as nodes, and their 4 constituent 3-bubble intersection
loci as connections, it is clear that we are generating a graph
with 4-fold connectivity. Note also that the precise location of
the strings on this graph is determined by the particular phase
realisation, enabling us to use the same graph to generate many
different string sets.

This algorithm is the natural extension of that employed in 2
dimensions to assess the effect of bubble wall speed on vortex
formation \cite{BKVV}. Its 3 dimensional realisation, including
the formation of both strings and monopoles, is described in
full detail elsewhere \cite{B}.

\section{Implemention}

This algorithm is clearly much more complex (and computationally
intensive) than its lattice-based counterparts; indeed the
limiting factor in its application will time and again prove to
be its CPU-time.

In a finite simulation we can never hope to represent the
longest strings --- for a string brearing field with correlation
length $\xi$ in a box of side $L$ any string longer than
$O(L^{2}/\xi)$ is likely to cross the boundary. If we leave the
boundaries open we can estimate the open string energy density
by seeing whether the number density of boundary crossings
reaches a non-zero asymptote as the box size increases
\cite{VV}. If instead we impose periodic boundaries, forcing all
strings to be loops, we can use the theoretical scarcity of long
closed loops to identify any excess as representing the open
strings.  We adopt the second approach for a number of reasons.
Since the hypothesis under consideration is that the string
energy density is due to loops alone we are looking for exactly
the excess that might occur in a periodic simulation.  Moreover,
the open boundary approach requires runs on a wide range of box
sizes to test for an asymptotic limit, and hence falls foul of
our CPU-time constraints. Finally the requirement that all
strings be closed provides a useful check that the phase
transition has completed and that the bubbles have indeed filled
the entire simulation volume by the end of the run, as well as
an overall test of the robustness of the code.

Since we wish to calculate many-bubble intersection points
periodicity is implemented by incorporating 26 copies of the
simulation volume to surround it. For a simulation of box size
$L$ and duration $T$ we then consider only those bubbles within
$T$ of the central volume, and only those intersections that
occur inside it. We therefore have to balance the wish to
maximise the ratio $L/T$ so as to minimise the periodic copies,
and the increased number of bubbles necessary to fill a larger
volume in a shorter time. Again the final constraint is
CPU-time. Note that the simulation size and time are
dimensionless parameters and that the physical scale of the
system is set by the bubble nucleation rate, and hence by
varying the number of bubbles nucleated. 

\section{Results}

We determine the graphs associated with 5 simulations each
starting with 5000 bubbles in a periodic box of size $5^{3}$
with a simulation runtime of 1 in units of the box length, each
graph requiring approximately 250 hours of CPU time on a
twin-processor Sun SPARC 10. We then generate the associated
string populations from 1000 different random phase realisations
in each case, giving some $N_{\rm s} \sim 300000$ strings.
Figure 1 shows the normalised number density $N(l)$ of all these
strings binned by length against the length $l$.

We can immediately see that there are three phases present:
\begin{itemize}
\item A low-end tail population of loops with $l < l_{\rm o}$,
with $l_{\rm o} \sim e^{0.5}$ here. Clearly the loop
distribution cannot follow equation [\ref{eLD}] down to
arbitrary short lengths, since this would generate an infinite
energy density. This population has never been explicitly
identified before, since lattice-based simulations automatically
impose a low-end cutoff of order a few lattice spacings; it is
worth noting that in this simulation the range of loop lengths
covers more than 6 orders of magnitude.
\item A population of scale-invariant Brownian loops with number
density falling as $l^{-5/2}$.
\item A high-end tail population of loops with $l_{\rm 1} < l <
l_{\rm max}$, with $l_{\rm 1} \sim e^{3.5}$ here (note that
$e^{3.5} \sim 5^{2}$, as expected). The truncation of the
longest strings (both long finite closed loops and infinite open
strings) by the periodic boundaries generates an excess of
strings longer than $l_{\rm 1}$, so that in this region the loop
number density falls more slowly, as $l^{-\alpha}$ with $\alpha
< -5/2$.
\end{itemize}

The final feature to note is that since we have a finite total
number of loops $N_{\rm s}$ in our simulation there is a minimum
mesurable non-zero normalised bin number density of $1/N_{\rm
s}$, clearly identifiable as the extreme high-end plateau in
figure 1. As the string number density approaches this the
relative error in the measurements can be seen to increase
dramatically. These extreme high-end points then correspond to
those bins which happen to contain one or more strings, despite
having a theoretical $\delta n(l) < 1/N_{\rm s}$.

This is in many ways a familiar picture, quantitatively giving
the predicted behaviour of the mid-range loops and at least
qualitatively the deviations at either extreme. The first
indication of significant differences comes with the value of
$\alpha$. Analytic estimates for random walks on a periodic
cubic lattice give $\alpha = 1$ \cite{ACR}, whereas we find
$\alpha \sim 1.9$. The fact that our long loop distribution
falls off much faster than a random walk model would predict
indicates that the amount of long string is much less than in
such models.

The periodicity of our simulation forces any open string back
into the box, where it would appear in the high end tail. We can
now estimate its energy density as the difference between the
observed energy density in all strings longer than $l_{\rm 1}$,
$E^{\rm obs}(l \geq l_{\rm 1})$, and the theoretical energy
density in loops longer than $l_{\rm 1}$ alone,
\begin{eqnarray}
E^{\rm th}_{\rm l}(l \geq l_{\rm 1}) & = & 
k \int^{\infty}_{l_{\rm 1}} l^{-3/2} {\rm d}l \nonumber \\
                   & = & 2 \; k \; l_{\rm 1}^{-1/2}
\end{eqnarray}

To determine the normalisation $k$ we can use the convergence of
the observed and theoretical loop distributions in the region
$l_{\rm o} \leq l \leq l_{\rm 1}$, giving
\begin{equation}
k = \frac{E^{\rm obs}(l_{\rm o} \leq l \leq l_{\rm 1})}
{2 (l_{\rm o}^{-1/2} - l_{\rm 1}^{-1/2})}
\end{equation}

The remarkable result when we do so for each of the 5 runs is
that
\begin{equation}
f_{\rm o} = \frac{E^{\rm obs}(l \geq l_{\rm 1}) - E^{\rm
th}_{l}(l \geq l_{\rm 1})}{E^{\rm obs}(l \geq 0)} = 0.0062
\end{equation}
(with a variance of 0.0050) compared to the $0.66 - 0.8$ we
might expect. We can account for {\em all} the energy density in
the strings as being due to a single population of
scale-invariant Brownian loops.

\section{Conclusion}

We have used a minimal dynamical model of a first-order phase
transition to generate an initial string distribution whose
statistics are significantly different from the standard
lore. We find that the loops short enough to be unaffected by
the box size are scale-invariant and Brownian (except at the
very shortest lengths) as before, but that now the energy
density in the longer loops can also be accounted for simply by
this distribution continued out to infinity. The absence of an
energy excess in the long loops means that we find no need to
include a second population of open strings. Moreover we can
readily see why all lattice-based work necessarily generates
open string, whether or not it should be present, whilst
graph-based work need not. Intriguingly, however, recent lattice
simulations show that it is also possible to achieve a very
significant reduction in the fraction of open string either by
increasing the variance of the sizes of the initial phase
domains \cite{KY}, or by decreasing the index of the power
spectrum of the string-bearing field \cite{V,RY}.

Our simulations are certainly constrained by their CPU-time
demands. The range of the central, scale-invariant Brownian
loop, distribution is narrower than we might like. However this
is an extremely difficult problem to address; the number of
bubbles required to fill the simulation increases with the
volume $L^{3}$, and the number of 4-nodes to calculate then
increases as $N^{4}$. Since the runs presented here are already
taking around 250 hours, even an order of magnitude increase in
$L^{2}$ would be prohibitively expensive.

We can expect the consequences of our results for string
cosmologies to be profound. Simulations of string evolution with
all the open string removed suggest that such distributions may
reach a different scaling solution from the usual one \cite{F}.
Furthermore, current models of string-induced density
perturbations focus on the wakes of long strings, and if we
remove the open string such models will require a much higher
string number density to have sufficient large loops to generate
the required perturbations. However this number density is
independently bounded from above by the millisecond pulsar
limits on gravitational wave production in the decay of small
loops. Whether cosmic string scenarios can be made to fit these
new constraints remains an open question.

\acknowledgements

The author acknowledges the support of NSF grant PHY-9453431,
and wishes to thank Andy Albrecht, Ed Copeland, Pedro Ferreira,
Mark Hindmarsh, Tom Kibble, James Robinson, Paul Shellard, Karl
Strobl, Alex Vilenkin and Andrew Yates for useful discussions.

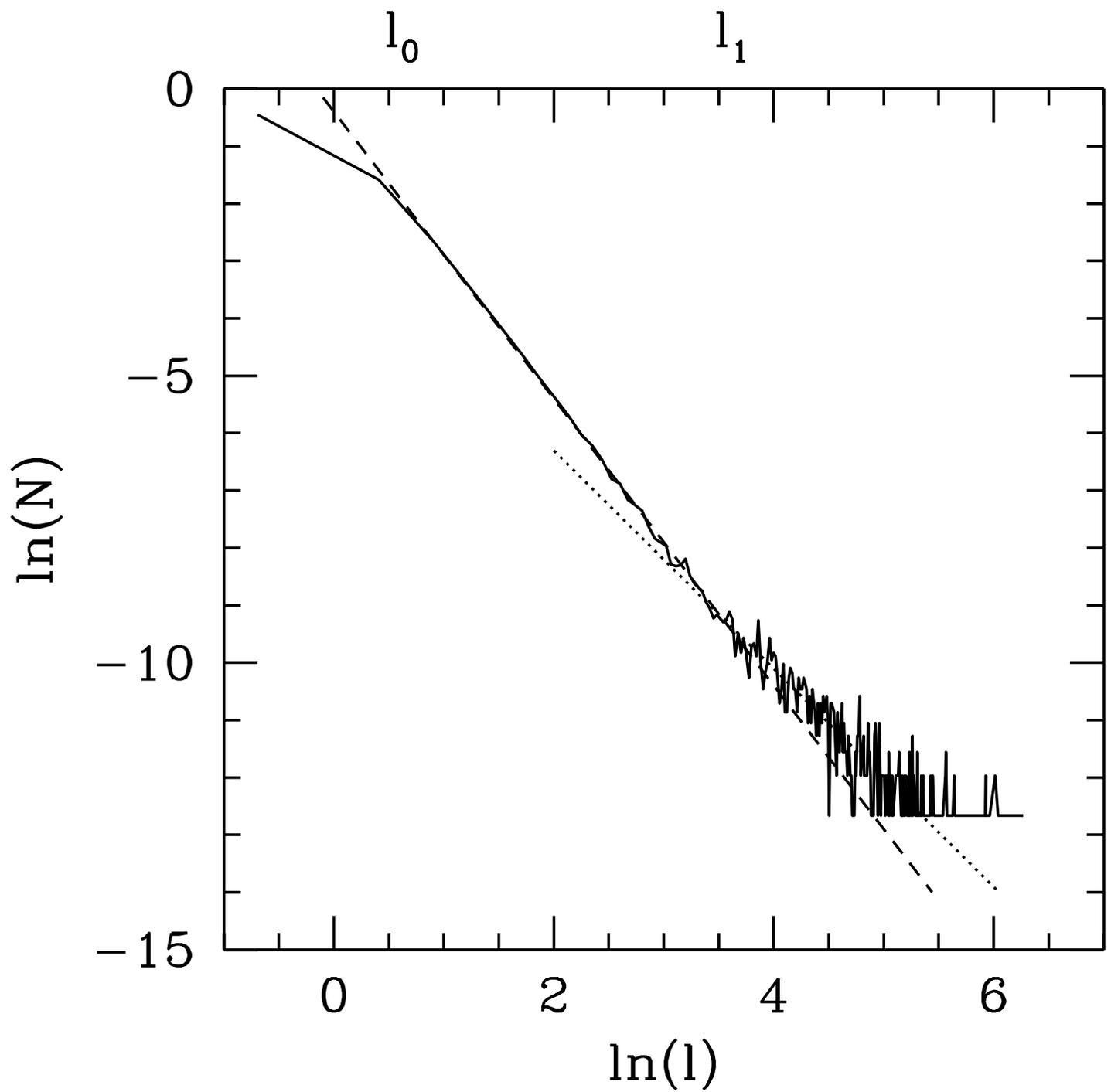
\begin{figure}

\caption{}{A log-log plot of the normalised number density
$N(l)$ of strings binned by length against their length $l$. The
solid line is the simulation data, the dashed line the best fit
to the Brownian regime, $N(l) = 0.67 \; l^{-2.5}$, and the
dotted line the best fit to the periodicity-distorted regime,
$N(l) = 0.082 \; l^{-1.9}$.}

\end{figure}

\end{document}